\documentclass[conference]{IEEEtran}
\IEEEoverridecommandlockouts
\usepackage{array}
\newcolumntype{P}[1]{>{\centering\arraybackslash}p{#1}}
\usepackage{multirow}
\usepackage{cite}
\usepackage{makecell}
\usepackage{algorithm}
\usepackage{algpseudocode}
\usepackage{tabularx}  % 放在 preamble，一次即可
\usepackage{amsmath,amssymb,amsfonts}
\usepackage{graphicx}
\usepackage{textcomp}
\usepackage[dvipsnames]{xcolor}
\usepackage{xcolor}
\usepackage[hidelinks]{hyperref}

\usepackage{array,makecell,ragged2e,booktabs,makecell}
\newcolumntype{L}[1]{>{\RaggedRight\arraybackslash}p{#1}}
\newcolumntype{C}[1]{>{\centering\arraybackslash}p{#1}}
\usepackage{array}

\def\BibTeX{{\rm B\kern-.05em{\sc i\kern-.025em b}\kern-.08em
    T\kern-.1667em\lower.7ex\hbox{E}\kern-.125emX}}
\begin{document}

\title{Scam Shield: Multi-Model Voting and Fine-Tuned LLMs Against Adversarial Attacks }

\author{\IEEEauthorblockN{Chen-Wei Chang\IEEEauthorrefmark{1},
Shailik Sarkar\IEEEauthorrefmark{1},
Hossein Salemi\IEEEauthorrefmark{2},
Hyungmin Kim\IEEEauthorrefmark{1},
Shutonu Mitra\IEEEauthorrefmark{1},\\
Hemant Purohit\IEEEauthorrefmark{2},
Fengxiu Zhang\IEEEauthorrefmark{3},
Michin Hong\IEEEauthorrefmark{4},
Jin-Hee Cho\IEEEauthorrefmark{1},
Chang-Tien Lu\IEEEauthorrefmark{1},
}
\IEEEauthorblockA{\IEEEauthorrefmark{1} Department of Computer Science, Virginia Tech, USA}% <-this % stops an unwanted space
\IEEEauthorblockA{\IEEEauthorrefmark{2}Department of Information Sciences and Technology,
\IEEEauthorrefmark{3}School of Policy and Government, George Mason University, USA}
\IEEEauthorblockA{\IEEEauthorrefmark{4}School of Social Work, Indiana University, USA}
}
\maketitle

\begin{abstract}
Scam detection remains a critical challenge in cybersecurity as adversaries craft messages that evade automated filters. We propose a \textit{Hierarchical Scam Detection System} (HSDS) that combines a lightweight multi-model voting front end with a fine-tuned LLaMA~3.1~8B Instruct back end to improve accuracy and robustness against adversarial attacks. An ensemble of four classifiers provides preliminary predictions through majority vote, and ambiguous cases are escalated to the fine-tuned model, which is optimized with adversarial training to reduce misclassification. Experiments show that this hierarchical design both improves adversarial scam detection and shortens inference time by routing most cases away from the LLM, outperforming traditional machine-learning baselines and proprietary LLM baselines. The findings highlight the effectiveness of a hybrid voting mechanism and adversarial fine-tuning in fortifying LLMs against evolving scam tactics, enhancing the resilience of automated scam detection systems.
\end{abstract}

\begin{figure*}[t]
    \centering
    \includegraphics[width=\textwidth]{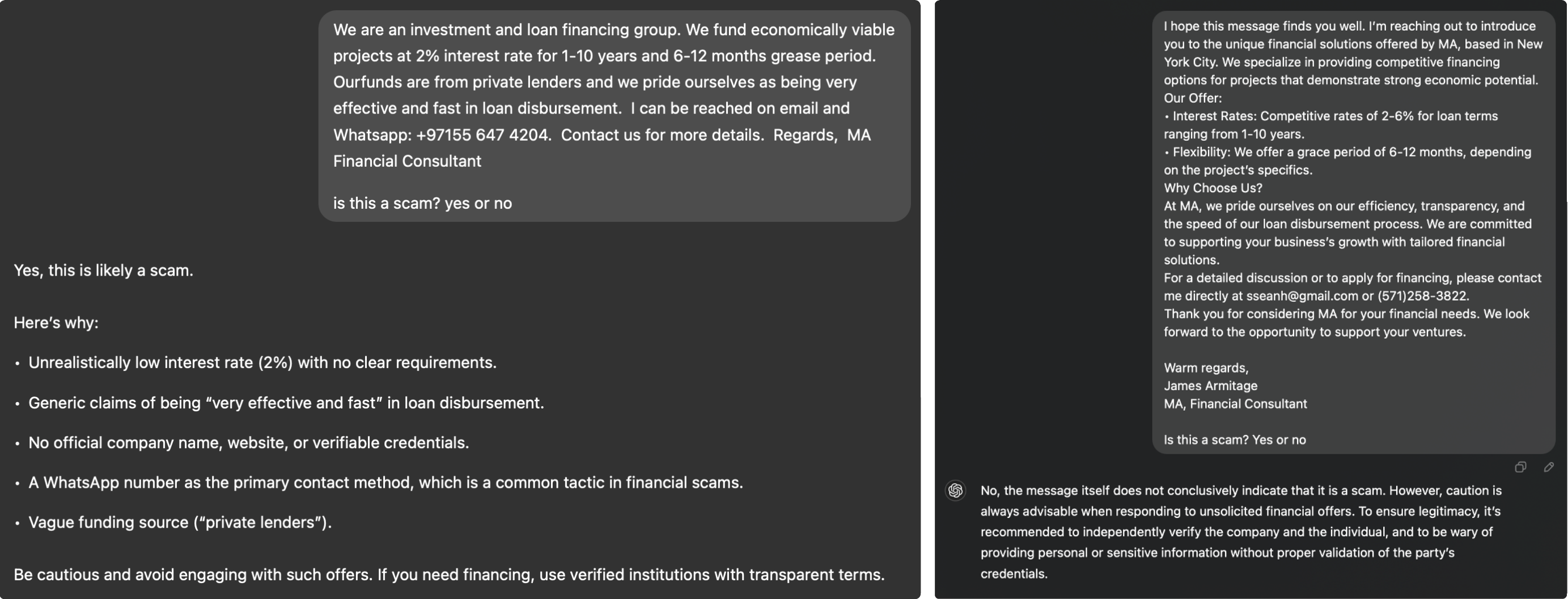} 
    
    \caption{Comparison of Original and Adversarially Modified Scam Messages with ChatGPT’s Scam Detection Response.}
    \label{fig:demo}
    \vspace{-3mm}
\end{figure*}
\section{Introduction}

Scam detection~\cite{shen2024combating} has become an increasingly critical challenge due to the rapid evolution of fraudulent tactics that exploit vulnerabilities in automated detection systems. Traditional approaches, including rule-based methods and machine learning models, have demonstrated limitations in adapting to novel and adversarially modified scam messages~\cite{salman2022empirical}. While Large Language Models (LLMs) have shown promise in improving scam classification accuracy, they remain susceptible to adversarial attacks~\cite{chang2024exposing}, such as subtle linguistic modifications by synonym substitution (e.g., replacing ``urgent'' with ``important''), sentence reordering, and adding harmless words to a scam message to hide warning signs without changing its main trick. As shown in \autoref{fig:demo}, where subtle linguistic modifications significantly impact detection performance.

Recent studies~\cite{chang2024exposing} investigated these vulnerabilities by curating datasets that include both original and adversarially modified scam messages to analyze how current LLMs fail under adversarial settings. While few-shot prompting with adversarially augmented data can improve robustness~\cite{rebuffi2021data}, models such as GPT\mbox{-}3.5 Turbo and Claude 3 Haiku still exhibit notable misclassification rates. Furthermore, smaller open-source models such as Meta LLaMA~3.1~8B Instruct perform poorly against adversarial examples, highlighting the need for more efficient fine-tuning strategies. The literature also lacks benchmarks for comprehensive adversarial scam detection. (See Definition~\ref{def:adv} for our operationalization of ``adversarial scam''.)  For brevity, we refer to Meta LLaMA~3.1~8B Instruct as LLaMA~8B throughout.

Therefore, we propose a Hierarchical Scam Detection System (HSDS) that integrates multi-model voting with a fine-tuned LLaMA 8B Instruct model to improve detection performance. Prior research has investigated ensemble-based classifiers~\cite{polikar2006ensemble} and employed proprietary LLMs for scam detection~\cite{chang2024site,jiang2024detecting}, but these approaches face several limitations. Ensemble-only methods lack robustness to adversarial perturbations because they rely on hand crafted features and shallow text patterns. Proprietary LLMs are closed-source and costly to deploy yet remain vulnerable while open-source LLMs (e.g., LLaMA~8B) offer a promising alternative for transparent and reproducible research, yet they typically require resource-intensive full-parameter fine-tuning, and their robustness against adversarial scams has not been systematically evaluated.

\begin{figure*}[t]
    \centering
    \includegraphics[width=\textwidth]{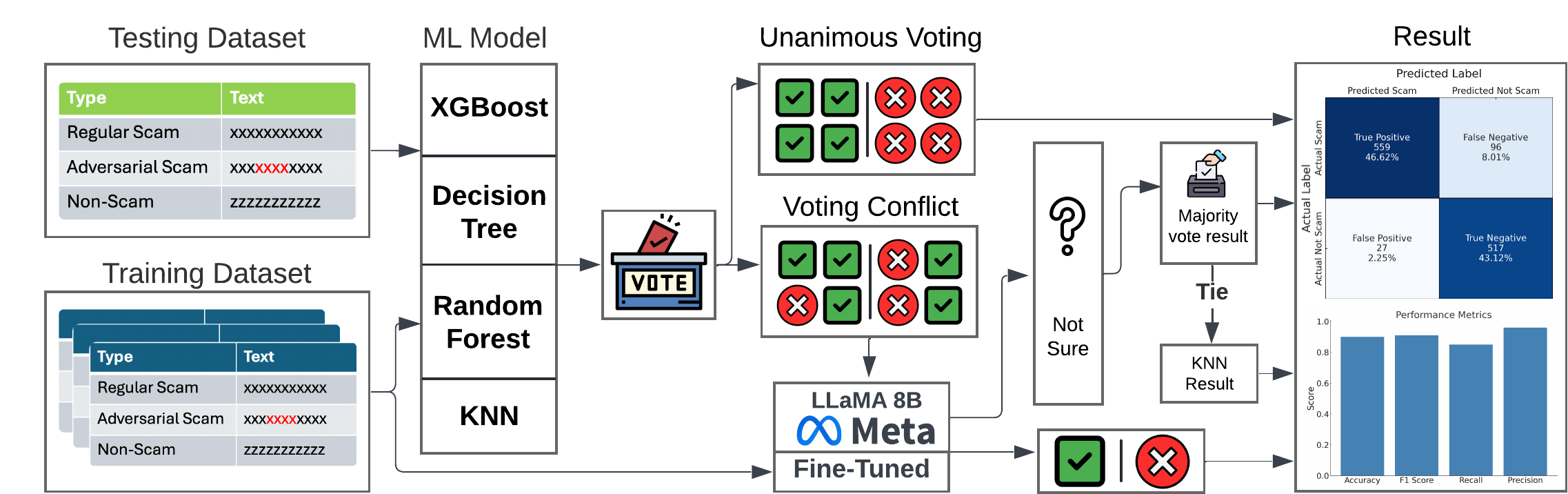} 
    
    \caption{Overview of our hierarchical scam detection system integrating multi-model voting and fine-tuned LLaMA 8B Instruct for enhanced detection accuracy.}
    \label{fig:methodology}
    
\end{figure*}

These gaps highlight the need for a framework that combines adversarial robustness, computational efficiency, and reproducibility on open-source platforms. To address this, our system makes the following key contributions:
\begin{itemize}
    \item \textbf{Hierarchical Scam Detection System with Multi-Model Voting:}
    We propose a hierarchical framework \autoref{fig:methodology} that integrates a multi-model ensemble with a fine-tuned LLaMA 8B Instruct model. This design aims to enhance robustness against both regular and adversarial scam messages while maintaining computational efficiency.

    \item \textbf{Developing a Novel Multi-Model Voting Mechanism:}
    To further improve accuracy, we employ an ensemble of four classifiers that provide initial predictions. If these models disagree, the final classification is handled by the fine-tuned LLaMA 8B Instruct model, ensuring higher confidence and reduced misclassification under adversarial perturbations.

    \item \textbf{Domain-Specific Fine-Tuning with LoRA on LLaMA 8B Instruct:}
    Instead of conventional full fine-tuning, we leverage Low-Rank Adaptation (LoRA) \cite{hu2021loralowrankadaptationlarge} to adapt LLaMA 8B Instruct. This approach significantly reduces computational overhead while achieving better adversarial detection accuracy than GPT-3.5 Turbo and Claude 3 Haiku.

    \item \textbf{Curating a Comprehensive Scam Dataset for Fine-Tuning}~\cite{trainingdataset}:
    We construct a 20{,}000-sample dataset enriched with adversarial scam messages, regular scam messages, and non-scam messages, applying data augmentation to foster diversity and improve model resilience.

    \item \textbf{Comprehensive Benchmarking of LLMs for Adversarial and Regular Scam Detection:}
    We systematically evaluate the robustness of various LLMs, examining their susceptibility to adversarial examples and identifying detection gaps. Our analysis includes a comparative study of different LLMs and explores how scam message categories affect classification accuracy.
\end{itemize}

Our results demonstrate that a multi-model voting approach improves robustness against adversarial scams, presents a pragmatic solution for real-world deployment, strategically mitigating the high inference latency and cost of LLMs by reserving their powerful analytical capabilities for only the most ambiguous cases, thereby achieving a robust and efficient defense against adversarial threats.

\section{Related Work}

\paragraph{\bf Overview of Existing Approaches.} 
Early scam detection relied on rule-based methods and machine learning models, such as Decision Trees~\cite{calderon2020filipino}, Support Vector Machines (SVMs), and Random Forests~\cite{dileep2021novel}, which leveraged handcrafted features to identify fraudulent patterns. More advanced models, such as XGBoost and K-Nearest Neighbors (KNN)~\cite{kannagi2023intelligent}, were later introduced to improve performance.

Recent advancements in Large Language Models (LLMs), such as GPT-3.5 Turbo and Claude 3 Haiku, have demonstrated strong zero-shot and few-shot learning capabilities for scam detection~\cite{jiang2024detecting}. To further enhance their performance, methods such as adversarial training, data augmentation, and fine-tuning have been explored~\cite{christophe2024med42}. LoRA (Low-Rank Adaptation)~\cite{hu2021loralowrankadaptationlarge} has emerged as an efficient fine-tuning alternative that reduces resource requirements.

Ensemble learning has been widely adopted in text classification to enhance robustness~\cite{gautam2023email}. We propose a hierarchical multi-model voting system that integrates traditional classifiers with a fine-tuned LLaMA 8B Instruct model for final decisions. Our approach aims to balance accuracy and computational efficiency while improving adversarial scam detection performance.

\paragraph{\bf Limitations of the State-of-the-Art.} 
Despite their effectiveness, existing approaches face several limitations. Traditional machine learning models, such as Decision Trees~\cite{calderon2020filipino}, SVMs, and Random Forests~\cite{dileep2021novel}, rely on handcrafted feature extraction, making them inflexible to evolving scam patterns. These methods are also highly susceptible to adversarial modifications, which significantly degrade their performance~\cite{gao2018black}. More advanced models like XGBoost and KNN~\cite{kannagi2023intelligent} offer improved performance but still lack the adaptability required to handle emerging scam variations effectively.

LLMs such as GPT-3.5 Turbo and Claude 3 Haiku have demonstrated strong zero-shot and few-shot learning capabilities in scam detection~\cite{jiang2024detecting}. However, these models remain highly vulnerable to adversarial text perturbations, which can lead to high misclassification rates in security-critical applications~\cite{chang2024exposing}. To mitigate such vulnerabilities, adversarial training and fine-tuning have been explored~\cite{christophe2024med42}. However, full fine-tuning is computationally expensive and resource-intensive. LoRA has been proposed as a more efficient fine-tuning approach~\cite{hu2021loralowrankadaptationlarge}, but its effectiveness in adversarial settings remains an open question.

While ensemble learning has been widely applied in text classification tasks~\cite{gautam2023email}, its potential for adversarial scam detection remains largely underutilized. Most existing methods focus on improving robustness at an individual model level, rather than leveraging ensemble techniques to enhance detection accuracy and resilience against adversarial attacks. These limitations motivate our proposed approach, which integrates fine-tuned LLMs with ensemble learning to enhance adversarial scam detection while maintaining computational efficiency. While newer proprietary LLMs such as GPT-4o and Claude 3.5 Sonnet have recently demonstrated stronger overall performance, systematic evaluation of their robustness under adversarial scam detection remains scarce. Due to computational and resource constraints, we adopt GPT-3.5 Turbo and Claude 3 Haiku as representative baselines. These models are widely used in prior studies and thus provide a consistent and comparable benchmark for evaluating our proposed approach.

\section{System Design and Methodology}

\subsection{Dataset for Fine-Tuning and Augmentation}
To enhance scam detection accuracy and resilience against adversarial attacks, we curated a comprehensive scam dataset \autoref{tab:dataset-example} by collecting 5,000 messages from \textit{Kaggle}~\cite{fraudulent_job_posting_2023, email_spam_2023, spam_or_not_spam_2023} and \textit{ScamWarners}\cite{scamwarners_website}, expanding it to 20,000 samples using data augmentation~\cite{maharana2022review}. The dataset consists of adversarial, regular, and non-scam messages to ensure balanced distribution for robust learning.

\noindent\textbf{Operational Definition.}
\label{def:adv}
We consider adversarial scam variants generated by a bounded set of text transformations (synonym replacement, random deletion with probability 10\% per token, and sentence reordering) that preserve meaning and readability while aiming to reduce detector confidence under a text-only, black-box attacker.

We applied three augmentation methods to improve adversarial robustness: 
\begin{itemize}
    \item \textbf{Synonym Replacement}: Words are substituted with their synonyms from \textit{WordNet} while preserving meaning.
    \item \textbf{Random Deletion}: Words are randomly removed with a probability of 10\% to introduce textual variations.
    \item \textbf{Sentence Shuffling}: The order of sentences is randomized to maintain semantic coherence while modifying structure.
\end{itemize}
By introducing linguistic diversity and complexity \cite{miestamo2017linguistic},
we ensure that the dataset remains diverse and challenging for detection models, improving their ability to recognize sophisticated scams.

\subsection{Generation of Adversarial Scam Messages}
To generate adversarial scam messages for augmentation, we employed a prompt engineering strategy. The process is automated by a program that acts as an assistant, rephrasing scam messages for research purposes. The assistant follows these guidelines for all rewrites:

\begin{enumerate}
    \item \textbf{Format:} Generate only the message content without any subject line or email header.
    \item \textbf{Remove Obvious Scam Indicators:} Avoid urgent requests or unusual payment demands.
    \item \textbf{Adjust Tone:} Use a professional, neutral tone throughout the message.
    \item \textbf{Retain Key Content:} Preserve the core information of the scam while rephrasing it in a more legitimate manner.
    \item \textbf{Add Limited Credibility:} Include general references to known locations or institutions sparingly to subtly enhance credibility.
    \item \textbf{Avoid Placeholders:} Do not use placeholder text such as \verb+[Your Name]+, \verb+[Position]+, \verb+[Contact]+, \verb+[Company]+, or \verb+[Location]+. Use specific details instead.
\end{enumerate}

\begin{table}[h]
\centering
\caption{\small Examples of Dataset for Fine-Tuning}
\label{tab:dataset-example}
\renewcommand{\arraystretch}{1.1}
\small
\setlength{\tabcolsep}{3pt}
\begin{tabular}{|P{2.9cm}|p{5.6cm}|}
\hline
\textbf{Label} & \textbf{Text} \\
\hline
Original recruitment scam message & Sir/Madam: 

Qualified and willing to be our USA customers Representative, collections and book-keeping? This offer is with a monthly salary of \$5,000 and benefits. (Part-time Job) ends 30 December 2020. Reply for more details, contact Mr. Martin Hendy AT: martinshendron@gmail.com 

--Best Regards. 

Miss. Blessing Alah Business Coordinator Well and Able Holdings 

Pte Ltd No.23 Genting Road \#03-01 

Chevalier House Singapore 349481. 

www.wellnable.com \\
\hline
Adversarial recruitment scam message & Hello,

We are looking for a part-time representative in the United States to assist with client coordination, collections, and bookkeeping. The role offers \$5,000 monthly compensation plus benefits, with flexible hours.

For further details, please contact Mr. Martin Hendy at: martinshendron@gmail.com.

Best regards,

Miss Blessing Alah

Business Coordinator

Well and Able Holdings Pte Ltd

23 Genting Road, \#03-01, Chevalier House

Singapore 349481

www.wellnable.com \\
\hline
Non-scam recruitment message & Hey Ashley,
Can you please confirm your availability for the meeting next week? We need to finalize the schedule. 

Cheers,
Tracie Gutierrez \\
\hline
\end{tabular}
\end{table}
\subsection{Experimental Methodologies}

Our hierarchical scam detection system consists of a multi-model voting framework integrated with a fine-tuned \textbf{LLaMA 8B Instruct} model, designed specifically to improve robustness against adversarial scam messages.  We selected the LLaMA 8B Instruct as an example of an LLM that performs poorly on adversarial scam messages~\cite{chang2024exposing}, intending to enhance its performance through fine-tuning and the proposed multi-model voting technique. The detailed workflow of our approach is illustrated in Figure~\ref{fig:methodology}.

\begin{itemize}
    \item \textbf{Dataset Preparation}: We use distinct datasets for the training and testing phases:
    \begin{itemize}
        \item \textbf{Training Dataset}: Comprises 20,000 messages evenly categorized into three classes: \textit{regular scam, adversarial scam, and non-scam messages}. This balanced dataset enables comprehensive learning and robust generalization.
        \item \textbf{Testing Dataset}: Utilizes an existing dataset~\cite{chang2024exposing},which includes both original and adversarially modified examples, ensuring consistency for a precise comparative evaluation of adversarial scam detection performance.
    \end{itemize}

    \item \textbf{Multi-Model Voting}: We implement an ensemble voting mechanism involving four machine learning classifiers: \textit{XGBoost, Decision Tree, Random Forest, and K-Nearest Neighbors (KNN)}. Each model independently classifies incoming messages, and the preliminary classification decision is established through majority voting across these models.

    \item \textbf{Handling Uncertain Cases}: If the initial voting outcomes are inconclusive (no majority consensus), the system forwards the message to a specialized fine-tuned LLaMA model for further classification.

    \item \textbf{Fine-Tuned LLaMA 8B Instruct for Final Decision}: The LLaMA 8B Instruct model is fine-tuned with the \textit{LoRA} method to enhance performance while conserving computational resources. This fine-tuned model acts as the decisive classifier when initial voting outcomes from the four ML models are uncertain. If the LLaMA 8B Instruct model returns an uncertain response, we trigger the fallback path defined in Algorithm~\ref{alg:voting_system}: an ensemble majority vote among \{RF, DT, XGB, KNN\}; if this vote ties, we default to KNN as the final tie-breaker.

    \item \textbf{Performance Evaluation}: System performance is rigorously evaluated using standard classification metrics, including \textit{accuracy, F1-score, recall, and precision}.
\end{itemize}

\autoref{fig:methodology}  provides overview of our hierarchical scam detection pipeline, showing the interaction between the dataset, machine learning models, and the fine-tuned LLaMA 8B Instruct decision-making process.

\subsection{Fine-Tuning Method}

To enhance the robustness of scam detection while minimizing computational costs, we fine-tuned LLaMA 8B Instruct using \textit{LoRA} (Low-Rank Adaptation)~\cite{hu2021loralowrankadaptationlarge}. LoRA enables efficient adaptation by injecting trainable low-rank matrices into transformer layers, significantly reducing memory and computation requirements compared to full fine-tuning. 

Using LoRA, we update only a small fraction of the model’s total parameters:
\begin{itemize}
    \item \textbf{Trainable Parameters} (\( P_t \)): 8,388,608
    \item \textbf{Total Parameters} (\( P_{\text{total}} \)): 8,037,076,992
    \item \textbf{Trainable Percentage} (\( P_t / P_{\text{total}} \)): 0.1\%
\end{itemize}

This significantly reduces computational overhead, enabling efficient fine-tuning on resource-limited hardware. We apply LoRA to the \(q_{\text{proj}}\) and \(v_{\text{proj}}\) projection matrices in each transformer layer; each targeted module introduces two low-rank matrices (A and B), i.e., \(2\,d_h r\) parameters per module.

\noindent\textbf{Parameter count.}
\begin{align}
    P_t &= 2_{\text{(A/B)}} \times 2_{\text{(q,v)}} \times d_h \times r \times L \\
        &= 4\, d_h r L \\
        &= 4 \times 4096 \times 16 \times 32 \\
        &= 8{,}388{,}608.
\end{align}

where \( d_h = 4096 \) is the hidden size of LLaMA 8B Instruct, \( r = 16 \) is the LoRA rank, and \( L = 32 \) is the number of transformer layers.

\begin{table}[t]
\caption{Hyperparameter and LoRA Configuration}
\label{tab:lora_hyperparams}
\centering
\footnotesize
\renewcommand{\arraystretch}{1.3}
\begin{tabular}{ll|ll}
\toprule
\textbf{Hyperparameters} & \textbf{Values} & \textbf{LoRA Config} & \textbf{Values} \\
\midrule
Learning Rate        & $3 \times 10^{-5}$     & Rank ($r$)        & 16 \\
Epochs               & 15                     & Alpha ($\alpha$)  & 32 \\
Batch Size           & 4                      & Dropout           & 0.05 \\
Grad. Accum. Steps   & 32                     & Target Modules    & $q_{\text{proj}}, v_{\text{proj}}$ \\
\bottomrule
\end{tabular}
\end{table}

\textbf{Additional Fine-Tuning Technical Details:}

\begin{itemize}
    \item \textbf{Quantization:} Employed 4-bit BitsAndBytes quantization to minimize memory usage while preserving model accuracy.
    \item \textbf{Instruction-Response Formatting:} Ensured consistency by aligning training data prompts with inference-time structured prompts.
    \item \textbf{Focused Training with Masking:} Loss computation exclusively targeted response tokens, ignoring instruction tokens, to improve training efficiency.
    \item \textbf{Post-Training Calibration:} Applied token probability adjustments to further enhance classification accuracy.
\end{itemize}

Experimental results indicate that our fine-tuned LLaMA 8B Instruct model \textbf{surpasses GPT-3.5 Turbo and Claude 3 Haiku} in adversarial scam detection, underscoring the effectiveness of adversarial training with LoRA-based fine-tuning.

\subsection{Multi-Model Voting System}

While traditional models like Random Forest (RF), Decision Tree (DT), XGBoost (XGB), and K-Nearest Neighbors (KNN) are individually less robust than our fine-tuned LLM in adversarial settings, they provide valuable and computationally efficient signals for initial scam detection. To balance efficiency and accuracy, we adopt a \textbf{hierarchical multi-model voting system} that integrates these classifiers with the fine-tuned LLaMA 8B Instruct.

Our hierarchical system uses four traditional models for rapid initial classification. Any disagreement among them escalates the decision to our fine-tuned LLaMA model. Should the LLM provide an uncertain response, a fallback mechanism is triggered: a majority vote is conducted among the traditional models, and if tied, the system defaults to the prediction from KNN. We selected KNN as the final tie-breaker because it achieved the strongest standalone performance in our evaluations . This layered approach enhances robustness while minimizing computational overhead.

The decision flow is prioritized as follows:
\begin{enumerate}
    \item \textbf{Unanimous Agreement:} If RF, DT, XGB, and KNN all produce the same prediction, that result is accepted.
    \item \textbf{Expert Adjudication:} If there is any disagreement, the message is passed to the fine-tuned LLaMA 8B Instruct for a more nuanced classification. Its decision is final.
    \item \textbf{Fallback Resolution:} If the LLM returns an uncertain result (i.e., its output is unparsable), a majority vote is taken from the four traditional models. If this vote results in a tie, the system defaults to the prediction from KNN.
\end{enumerate}

The voting process is formally detailed in Algorithm~\ref{alg:voting_system}.

\begin{algorithm}[h]
\caption{Multi-Model Voting System}
\label{alg:voting_system}
\small
\begin{algorithmic}[1]
\Require Message $M$, feature vector $X$
\State $preds \gets [\ ]$
\For{each $model \in \{RF, DT, XGB, KNN\}$}
    \State $preds.\text{append}(model.\text{predict}(X))$
\EndFor
\If{$|\text{set}(preds)| == 1$}
    \State \Return $preds[0]$ \Comment{All models agree}
\Else
    \State $llm\_result \gets \text{classify\_message\_llm}(M)$
    \If{$llm\_result == -1$} \Comment{LLM uncertain}
        \State $(vote,\ is\_tie) \gets \text{majority\_vote}(preds)$
        \If{$ \textbf{not}\  is\_tie$}
            \State \Return $vote$
        \Else
            \State \Return $KNN.\text{predict}(X)$ \Comment{Tie $\Rightarrow$ fallback to KNN}
        \EndIf
    \Else
        \State \Return $llm\_result$
    \EndIf
\EndIf
\end{algorithmic}
\end{algorithm}

\subsection{LLaMA-Based Scam Classification}

To effectively detect adversarial scam messages, we leverage our fine-tuned LLaMA 8B Instruct model. The model is prompted with a structured query format and classifies messages into \texttt{scam} or \texttt{not scam} using a binary response system.

The classification prompt follows a strict instruction:
\begin{quote}
\textit{``You are a scam‑detection assistant. Respond ONLY with 'yes' (scam) or 'no' (not scam)''}
\end{quote}

Given an input message \( M \), the model processes the text and generates a response in a zero-shot setting. The detection process is summarized in Algorithm~\ref{alg:llm_classification}.

\begin{algorithm}[h]
\caption{LLaMA 8B Instruct Scam Classification}
\label{alg:llm_classification}
\small
\begin{algorithmic}[1]
\Require Message $M$
\State Generate response $R$ using LLaMA 8B
\State Extract: $matches \gets$ \texttt{re.findall(\textquotesingle\textbackslash b(yes$\mid$no)\textbackslash b\textquotesingle, R)}
\If{$matches \neq \emptyset$}
    \If{$matches[-1] ==$ `yes`}
        \State \Return $1$ \Comment{Scam detected}
    \Else
        \State \Return $0$ \Comment{Not a scam}
    \EndIf
\Else
    \State \Return $-1$ \Comment{Uncertain response}
\EndIf
\end{algorithmic}
\end{algorithm}

\noindent\textbf{Decoding configuration.}
We use deterministic decoding to minimize variance: \texttt{temperature} = 0.0, \texttt{do\_sample} = \texttt{False}, \texttt{top\_k} = 1, and \texttt{max\_new\_tokens} = 10.
Under this setup, the model produces a short, stable answer for the binary decision.

\noindent\textbf{Decision and uncertainty rule.}
We use deterministic decoding and post-process by matching whole-word \texttt{yes}/\texttt{no}.
If at least one match appears, we take the last match as the decision (\texttt{yes}\(\to 1\), \texttt{no}\(\to 0\)).
If neither token appears, we return \(-1\) (uncertain) and follow the fallback path in Algorithm~\ref{alg:voting_system}. Perform an ensemble majority vote among \{RF, DT, XGB, KNN\}; if this vote ties, default to KNN.
This rule is deterministic under our decoding setup.

\section{EXPERIMENTAL RESULTS \& ANALYSES}
\paragraph{\bf Evaluation settings.}
We consider S1 Prompt-only (no fine-tuning). All models share the same few-shot template—and S2 Domain–fine-tuned (ours). LLaMA 8B Instruct is fine-tuned on in-domain data and evaluated \emph{zero-shot}. Proprietary models are not fine-tuned and thus are reported only under S1.

\paragraph{\bf Headline finding.}
In S2, our fine-tuned 8B (zero-shot) outperforms the 1 few-shot proprietary baselines on adversarial scams, supporting that \emph{training-time domain adaptation} can outperform \emph{inference-time prompting} in this 
setting.
\begin{table*}[t]
    \centering
    \caption{Adversarial scam benchmark under two settings:\\
    \textbf{S1 Prompted-only (no fine-tuning)} with few-shot ICL vs.
    \textbf{S2 Domain-finetuned (ours)} with zero-shot inference.}
    \label{tab:fine_tune_results}
    \small
    \begin{tabular}{llcccc}
    \toprule
    \textbf{Setting} & \textbf{Model} & \textbf{Accuracy} & \textbf{Precision} & \textbf{Recall} & \textbf{F1} \\
    \midrule
    \multirow{3}{*}{S1: Prompted-only (no FT), few-shot}
      & GPT-3.5 Turbo          & 0.78 & 0.85 & 0.77 & 0.82 \\
      & Claude 3 Haiku         & 0.70 & 0.66 & 0.82 & 0.73 \\
      & LLaMA-8B (no FT)       & 0.59 & 0.51 & 0.73 & 0.60 \\
    \midrule
    \multirow{2}{*}{S2: Domain-finetuned (ours), zero-shot}
      & LLaMA-8B (finetuned)              & \textbf{0.87} & \textbf{0.89} & \textbf{0.82} & \textbf{0.86} \\
      & LLaMA-8B (finetuned) + Voting     & \textbf{0.90} & \textbf{0.95} & \textbf{0.85} & \textbf{0.90} \\
    \bottomrule
    \end{tabular}

    \vspace{2mm}
    \footnotesize\emph{Fairness note:} S1 compares prompted-only models under the same few-shot template. S2 evaluates our domain-finetuned model in zero-shot (no in-context examples at inference). Proprietary models are not finetuned due to API constraints; thus S2 evidences a \textit{mechanism} comparison (training-time domain adaptation vs. inference-time prompting), not a like-for-like \emph{model} comparison.
\end{table*}
\subsection{Fine-Tuning Performance Evaluation on Adversarial Scam Detection}
We first compare our fine-tuned LLaMA-8B Instruct model against GPT-3.5 Turbo, Claude 3 Haiku, and LLaMA-8B using the adversarial scam dataset (Table~\ref{tab:fine_tune_results}). The fine-tuned model achieves an accuracy of 0.87, outperforming GPT-3.5 Turbo (0.78 few-shot) and significantly surpassing the baseline LLaMA-8B (0.59 few-shot). Unlike GPT-3.5 Turbo and Claude 3 Haiku, which require few-shot prompting to perform reasonably well, our model maintains high performance in a \textit{zero-shot setting}, underscoring its robustness without additional prompt engineering.

In terms of precision and recall, the fine-tuned LLaMA-8B achieves 0.89 and 0.82, respectively. By comparison, Claude 3 Haiku reaches a relatively high recall of 0.82 but a much lower precision of 0.66, resulting in frequent false positives. Our model offers a more balanced trade-off, detecting scams more accurately while reducing misclassifications. These improvements stem from adversarial training and LoRA-based efficient adaptation, which strengthen detection of deceptive text while remaining computationally manageable.

Overall, our domain-specific fine-tuning on adversarial data not only improves standard performance metrics but also increases resilience against evolving scam tactics. This adaptability is critical for real-world deployment, where adversarial spam and phishing strategies continue to evolve rapidly.

\subsection{Enhancing Scam Detection with Multi-Model Voting}

While previous studies have sometimes reported lower performance for traditional models such as Random Forest (RF), Decision Tree (DT), XGBoost (XGB), and K-Nearest Neighbors (KNN) in adversarial scenarios, our curated adversarial scam dataset shows that these models achieve reasonably strong results. Among them, KNN reached the highest performance (Accuracy 0.86, F1 0.87), followed by Random Forest (0.84 / 0.85), whereas Decision Tree and XGBoost performed somewhat lower (both around 0.75 / 0.77). Although none of these baselines surpassed our fine-tuned LLaMA 8B Instruct model, they provide complementary signals that we exploit through a multi-model voting system. This layered approach leverages their diverse strengths, yielding more robust detection while maintaining computational efficiency.

\subsection{Impact of Multi-Model Voting on Performance}
Next, we evaluated the benefit of the multi-model voting system by comparing the fine-tuned LLaMA-8B Instruct model both with and without voting (Table~\ref{tab:fine_tune_results}). When the voting mechanism is included, accuracy rises from 0.87 to 0.90, a gain of around 3.4\%, while precision improves from 0.89 to 0.95, marking an approximate 6.7\% increase. Recall also improves slightly from 0.82 to 0.85, and the F1-score increases from 0.86 to 0.90, representing a 4.6\% overall improvement.

These results demonstrate that the multi-model voting system effectively refines scam detection by leveraging complementary strengths from traditional machine learning models and LLM-based classification. By achieving higher accuracy, precision, and F1-score without sacrificing recall, the proposed approach ensures a balanced trade-off between robustness and efficiency, making it well-suited for real-world scam detection applications.

\subsection{Performance Evaluation on Additional Scam Datasets}
To further validate the robustness and generalizability of our fine-tuned LLaMA 8B Instruct model, which is trained on an adversarial scam dataset, we evaluated its performance ~\ref{table:different_datasets} on two additional, more general scam datasets. The model achieved an accuracy of \textbf{0.88} on \textbf{Dataset 2}~\cite{ds2} and \textbf{0.82} on \textbf{Dataset 3}~\cite{ds3}, demonstrating its strong ability to identify scam messages across diverse scenarios. These results reinforce the model’s suitability for broader real-world applications. 

\begin{table}[h]
\centering
\caption{Accuracy of Fine-Tuned LLaMA 8B Instruct Across Different Datasets}
\label{table:different_datasets}
\renewcommand{\arraystretch}{1.2}
\small
\begin{tabular}{|L{0.28\linewidth}|L{0.4\linewidth}|C{0.16\linewidth}|}
    \hline
    \textbf{LLM} & \textbf{Dataset Type} & \textbf{Accuracy} \\
    \hline
    \multirow{3}{*}{\centering\makecell{LLaMA 8B Instruct\\fine-tuned}}
        & Adversarial Scam         & 0.87 \\
    \cline{2-3}
        & General Scam (Dataset 2) & 0.88 \\
    \cline{2-3}
        & General Scam (Dataset 3) & 0.82 \\
    \hline
\end{tabular}
\end{table}

\subsection{Multi-Model Voting Performance on Different Scam Types}

\begin{table*}[t]
\centering
\footnotesize
\caption{Voting Performance on Different Scam Types}
\label{table:multi_model_performance}
\renewcommand{\arraystretch}{1.6}
\begin{tabularx}{\textwidth}{|X|X|X|X|X|X|X|X|}
    \hline
    & \textbf{Romance} & \textbf{Recruitment} & \textbf{Finance} & \textbf{Pet} & \textbf{Lottery} & \textbf{Loan}   \\
    \hline
    \textbf{Recall} & 0.7308 & 0.8370 & 0.9701 & 0.9722 & 0.7676 & 0.9789   \\
    \hline
    \textbf{F1 Score} & 0.8444 & 0.9112 & 0.9848 & 0.9859 & 0.8685 & 0.9894  \\
    \hline
\end{tabularx}
\end{table*}
To evaluate the effectiveness of our multi-model voting system across different scam categories, we analyzed its recall and F1 scores, as shown in Table~\ref{table:multi_model_performance}. The system performs exceptionally well in detecting \textit{finance}, \textit{pet}, and \textit{loan} scams, with recall scores above 0.97 and F1 scores above 0.98. These results demonstrate that scam types characterized by structured patterns or explicit transaction-related language (e.g., suspicious offers or payment requests) are consistently recognized. This also indicates that the voting mechanism effectively amplifies strong lexical or structural cues, leading to highly stable detection performance across these categories.

The strong outcomes in \textit{finance} and \textit{loan} scams in particular reflect the system’s ability to capture fraudulent financial practices, such as unusual payment terms or unrealistic lending conditions, while \textit{pet} scams are often identifiable through recurring signals like references to shipping charges or adoption fees. In these cases, the different components of HSDS play complementary roles: traditional classifiers efficiently capture repetitive surface patterns, whereas the fine-tuned LLaMA provides refined judgment when the text context is more nuanced. This layered process helps explain the consistently high performance observed for these categories.

However, \textit{romance} scams exhibit a comparatively lower recall of 0.7308, despite maintaining a fairly high F1 score of 0.8444. This discrepancy suggests that while the model can correctly identify many romance-related frauds, a subset of these messages remains difficult to classify. Unlike other scam categories, romance scams focus more on emotional manipulation rather than clear references to money or transactions, making them less straightforward for keyword-driven detection. \textit{Lottery} scams, with a recall of 0.7676 and an F1 score of 0.8685, also pose challenges due to their sometimes ambiguous language and reliance on promises of large winnings. These observations highlight that scams relying on implicit persuasion or ambiguous incentives still pose difficulties, and current classifiers may need further adaptation to capture these subtler patterns.

 As shown in Table~\ref{table:multi_model_performance}, it recalls money-related scams with high consistency, but has more difficulty capturing romance scams that rely on emotional language and distinguishing certain non-scam messages. Future steps such as more focused training, integrating semantic or context-aware features, and category-specific fine-tuning could further enhance its ability to detect diverse scam types.

\subsection{Evaluating Weighted Voting in Scam Detection}

\begin{table}[h]
\centering
\caption{Performance Comparison: Majority Voting vs. Weighted Voting}
\label{tab:weighted_voting_results}
\renewcommand{\arraystretch}{1.2}
\small
\begin{tabularx}{\linewidth}{|X|X|X|X|X|}
    \hline
    \textbf{Voting Method} & \textbf{Accuracy} & \textbf{Precision} & \textbf{Recall} & \textbf{F1 Score} \\
    \hline
    \textbf{Majority Voting} & 0.90 & 0.95 & 0.85 & 0.90 \\
    \hline
    \textbf{Weighted Voting} & 0.86 & 0.92 & 0.83 & 0.87 \\
    \hline
\end{tabularx}
\end{table}

\autoref{tab:weighted_voting_results} compares the performance of majority voting and weighted voting. In addition to majority voting, we explored a weighted strategy in which each model’s contribution is scaled by its individual performance. Random Forest and KNN were assigned a weight of 0.3 each due to their stronger results, while Decision Tree and XGBoost were given a weight of 0.2 each. The final decision is determined by the class with the highest cumulative weighted score.

The results show that majority voting achieves higher accuracy (0.90 vs. 0.86), precision (0.95 vs. 0.92), recall (0.85 vs. 0.83), and F1-score (0.90 vs. 0.87). This indicates that majority voting consistently outperforms the weighted approach under our current setup. Therefore, majority voting is the preferable choice for overall robustness, while weighted voting may still be explored in future work with alternative weighting schemes or domain-specific adjustments.

\subsection{Computational Efficiency Analysis}

Beyond improving classification accuracy, the proposed multi-model voting system significantly reduces computation time compared to direct inference using the fine-tuned LLaMA 8B Instruct model. We evaluated efficiency on a dataset containing 1200 messages, including adversarial scam, regular scam, and non-scam instances.

Table~\ref{tab:computation_time} shows the fine-tuned LLaMA 8B Instruct model takes 2296 seconds to process the full dataset, averaging 1.91 seconds per message. In contrast, the multi-model voting system significantly reduces inference time to 995 seconds (0.83 sec. per message), achieving a 56.7\% reduction in computational time. 

This evaluation was conducted on a MacBook Pro M2 Pro with 32GB RAM. The significant improvement in efficiency is attributed to the ability of traditional models (RF, DT, XGB, and KNN) to handle most classifications quickly, leveraging LLaMA only when necessary. This hybrid approach optimally balances computational cost and detection accuracy, making it a more practical solution for real-world scam detection deployments.

\begin{table}[t]
\caption{Computation Time Comparison: LLaMA 8B Instruct vs. Multi-Model Voting}
\label{tab:computation_time}
\centering
\footnotesize
\setlength{\tabcolsep}{4pt}
\renewcommand{\arraystretch}{1.5}
\begin{tabular*}{\linewidth}{@{\extracolsep{\fill}} lcc}
\toprule
\textbf{Method} & \textbf{Total Time (s)} & \textbf{Avg. Time per Message (s)} \\
\midrule
LLaMA 8B Instruct Fine-Tuned & 2296 & 1.91 \\
Multi-Model Voting  & 995  & 0.83 \\
\bottomrule
\end{tabular*}
\end{table}

\section{Conclusions and Future Work}

\noindent \textbf{Summary of Key Contributions.} In this work, we proposed a \textit{Hierarchical Scam Detection System} that integrates multi-model voting with a fine-tuned LLaMA 8B Instruct model. Our contributions include the creation of a comprehensive adversarial dataset, an efficient LoRA-based fine-tuning methodology, and the development of a novel voting framework. Collectively, these efforts advance scam detection by improving accuracy, adversarial resilience, and computational efficiency.

\paragraph{\bf Key Findings.}
Our experiments demonstrate that this hierarchical system significantly improves adversarial scam detection and reduces computational overhead, outperforming strong LLM baselines. The adversarial fine-tuning strategy with LoRA proved highly effective, creating a robust model that excels even in a zero-shot setting. We also confirmed that a majority voting rule surpassed a weighted approach for this task, and that traditional classifiers, while individually less potent, are crucial to the system's overall robustness and efficiency.

\paragraph{\bf Future Work.}
Future work should focus on benchmarking against state-of-the-art LLMs (e.g., GPT-4o) and developing specialized models for nuanced scams like romance fraud. Further enhancements include refining the voting mechanism with adaptive weighting and incorporating user feedback loops for continuous adaptation to evolving threats. Optimizing the architecture for real-time, high-throughput deployment remains a critical next step. Addressing these challenges will further enhance the effectiveness of AI-driven systems in combating sophisticated digital fraud.

Overall, this work highlights the significant potential of hybrid frameworks that combine fine-tuned large language models with ensemble voting to enhance adversarial scam detection. Future research addressing remaining gaps will further enhance real-world adaptability and effectiveness.

\section*{Acknowledgement}
This work is partly supported by the Commonwealth Cyber Initiative (CCI) through its Inclusion and Accessibility in Cybersecurity program.
\bibliographystyle{IEEEtran}
\bibliography{ref}

\vspace{12pt}

\end{document}